\documentclass[review]{elsarticle}

\usepackage{lineno,hyperref}
\modulolinenumbers[5]

\usepackage{graphicx}  
\usepackage{dcolumn}   
\usepackage{bm}        
\usepackage{amsmath}   
\usepackage{amssymb}   
\usepackage{amsfonts}
\usepackage{graphics}
\usepackage{epsfig}
\usepackage{units}
\usepackage[usenames]{color}
\graphicspath{{./img/}}
\usepackage{lineno}
\linenumbers
\modulolinenumbers[5]
\usepackage{fancyhdr}
\newcommand{\beq}{\begin{equation}}
\newcommand{\eeq}{\end{equation}}
\newcommand{\bqa}{\begin{eqnarray}}
\newcommand{\eqa}{\end{eqnarray}}

\begin{document}
\title{A Tale of Tails:  Photon Rates and Flow in Ultra-Relativistic Heavy Ion Collisions}
\author[bnl,ccnu]{Larry McLerran}
\author[bnl]{Bj\"orn Schenke}
\address[bnl]{Physics Dept, Bdg. 510A, Brookhaven National Laboratory, Upton, NY-11973, USA}
\address[ccnu]{Physics Dept, China Central Normal University, Wuhan, China}

\vspace{0.3cm}
\begin{abstract}
We consider the possibility that quark and gluon distributions in the medium created in high energy heavy ion collisions may be modified by a power law tail at energies much higher than the temperature.  We parametrize such a tail by Tsallis distributions with an exponent motivated by phenomenology.  These distributions are characterized by an effective temperature scale that we assume to evolve in time like the temperature for thermal distributions.  We find that including such a tail increases the rates for photon production and significantly delays the emission times for photons of a fixed energy.  We argue that these effects are sufficiently large that they should be able to account for photon yields and flow patterns seen in LHC and RHIC experiments. 
\end{abstract}

\maketitle
\section{Introduction}
Theoretical calculations of thermal photon emission from the quark gluon plasma \cite{Shuryak:1978ij,Chatterjee:2005de,Chatterjee:2008tp,Bratkovskaya:2008iq,Chatterjee:2011dw,Holopainen:2011pd,Dion:2011pp,vanHees:2011vb,Chatterjee:2013naa,Linnyk:2013hta,Shen:2013cca,Shen:2013vja} have difficulties describing the experimentally observed photon yield and elliptic flow in heavy ion collisions \cite{Adare:2008ab,Adare:2011zr,Wilde:2012wc,Lohner:2012ct}. Calculations result in photon transverse momentum spectra that are too shallow and approximately a factor of 4 smaller than the observed yield. Further, the photon elliptic flow $v_2$ is computed to be significantly lower than the observed $v_2$. The small computed photon $v_2$ is easily understood because photons are predominantly produced at early times (which correspond to high temperatures), when flow has not yet been built up. Various scenarios have been suggested in the literature to explain the discrepancy \cite{Liu:2009kta,Bzdak:2012fr,Muller:2013ila,Linnyk:2013hta,Linnyk:2013wma,Linnyk:2015tha,Monnai:2014kqa,
McLerran:2014hza,Monnai:2014qza,Gale:2014dfa}. 

In this paper we study the effect of deviations from thermal quark and gluon distributions in the plasma. In particular, we show that non-thermal tails, implemented by replacing Bose-Einstein and Fermi-Dirac distributions by the corresponding Tsallis distribution, have a significant effect on photon yields and typical emission times: As opposed to most other mechanisms that have been suggested to resolve the thermal photon puzzle, photon emission times are delayed and at the same time production rates are increased. This could resolve the puzzle by generating both increased photon yields and elliptic flow.

The paper is organized as follows:
We show that the photon yield in heavy ion collisions is sensitive to the high momentum region of the distribution in Section \ref{sec:review}, and demonstrate that enhanced tails can lead to both an increase of the photon yield and a delay of photon emission in Section \ref{sec:replace}. We then present a detailed quantitative calculation of the photon yield from a one-dimensionally expanding plasma with Tsallis quark and gluon distributions in Section \ref{sec:detail}. We conclude in Section \ref{sec:conc}.

\section{Review of Photon Production In 1+1D Hydrodynamical Models}\label{sec:review}
Photon production in heavy ion collisions is dominated by photons at a momentum scale $k \sim 4T$. 
To see this consider the formula for the rate of photon production from a thermalized
quark gluon plasma \cite{Baier:1991em,Kapusta:1991qp}
\begin{equation}\label{eq:thermalrate}
  E_\gamma \frac{dN_{\rm th}}{d^4x d^3p} = \frac{5}{9}{{\alpha_s \alpha} \over {2\pi^2}} T^2 e^{-E_\gamma/T} \ln\left(\frac{2.912}{4\pi\alpha_s}\frac{E_\gamma}{T}\right)\,,
\end{equation}
where $E_\gamma$ is the photon energy, $T$ the temperature of the medium, and $\alpha_s$ and $\alpha$ the strong and electromagnetic coupling, respectively.

To compute the photon yield from an expanding ideal gas in 1+1 dimensions, we integrate this rate over time using the time dependence of the temperature, which for an ideal gas of relativistic quarks and gluons is given by \cite{Shuryak:1978ij}
\begin{equation}
    t/t_0 = T_0^3/T^3\,.
\end{equation}
The four volume is 
\begin{equation}
  \int d^4x = A_T\int t\,dt
\end{equation}
where $A_T$ is the transverse area of the interaction region. Since 
\begin{equation}
  t\,dt  \propto {{ dT} \over T} {1 \over T^6}\,,
\end{equation}
upon inserting the rate at a given $T$ into the integral over the expanding system, we 
get 
\begin{equation}
  E_\gamma \frac{dN_{\rm th}}{d^3p} \propto \int \frac{dT}{T} \exp[-E_\gamma/T - 4 \ln(T/T_0)]\,.
\end{equation}
By determining the stationary point of the exponent we obtain the
typical emission energy  for photons to be
\begin{equation}
   E_\gamma \sim 4T
\end{equation}
up to logarithmic corrections.   (The best way to do the stationary phase distribution is in logarithmic coordinates, $\chi = \ln(T/T_0)$ where $dT/T = d\chi$.)

This emission occurs when we are in the tail of the exponential distribution.

\section{Replacing Thermal Distributions by Tsallis Distributions}\label{sec:replace}
Let us now suppose that the quark and gluon distributions were not purely Bose Einstein or Fermi Dirac distributions.
Since emission occurs at somewhat hard momenta, we might expect the distributions would be well approximated by an exponential with a power law tail.
An example which has this property is the Tsallis distribution,
\begin{equation}\label{eq:tsallis}
  f(E) = [1 + E/(aT)]^{-a}\,,
\end{equation}
where $E$ is the quark or gluon energy, and $a$ is a free parameter that determines the power law at large $E$.
For $E \ll aT$ this distribution is well approximated by an exponential, $ f \sim e^{-E/T}$, but for large energy it goes as $f \sim (E/aT)^{-a}$.  
For a thermal distribution undergoing 1+1 dimensional Bjorken expansion, the integral
\begin{equation}\label{eq:rho}
  \rho  = \int d^3p\, f \sim 1/t\,,
\end{equation}
as it should for a non-interacting gas.  
For proton+proton collisions, the measured distribution of produced charged particles is approximately a Tsallis distribution with $a \approx 6$.
In order that the Tsallis distribution and the thermal distribution describe the same number of particles, the temperature in the Tsallis distribution must be taken to be somewhat below that of the thermal distribution.  We will show this explicitly below.  We will see that this change does not affect the conclusion that the rate of photon emission goes up and the times of emission are lengthened for a Tsallis distribution relative to the thermal one.

In our analysis, we assume that the generalized (Tsallis) momentum distribution depends on only one variable, the temperature, i.e., the same principle scaling property as for thermal distribution functions. Such scaling behavior of distributions has been observed in simulations of the Glasma using classical field techniques \cite{Blaizot:2011xf,
Gelis:2013rba,Epelbaum:2014xea,Berges:2013eia,Berges:2013fga,Berges:2014yta}.  It is of course more complicated if the longitudinal momentum and transverse momentum scales evolve differently in time.  For simplicity, we assume only one scale, but the analysis could be extended.
It would of course be best to take the distribution directly derived from a first principle computation, and this might be possible in the future. 

Note that the distribution (\ref{eq:tsallis}) tends to 1 as $E \rightarrow 0$.  At high $p_T$ on the other hand, this distribution scales as $(T/E)^a$.  
The dependence of the multiplicity upon the number of participants at any time $t$ is
\begin{equation}
{{dN} \over {d^3p}} =A^{2/3} f\,,
\end{equation}
where $A$ is the number of participating nucleons. 
So for low energies, where $f\rightarrow 1$, the multiplicity distribution scales as the number of participants.  

However, at high energies, it scales as $A^{2/3} T^a$.  In saturation models, the initial temperature scales
like  $T \sim A^{1/6}$ so that for $a \sim 6$, we get a very rapid $A^{5/3}$ growth in the multiplicity.\footnote{To understand this dependence, note that the density of partons in the transverse plane $\rho \sim A^{1/3}$ determines the saturation 
  scale $Q_s^2 \sim \rho \sim A^{1/3}$.}

Note also that the low momentum part of the distribution does not evolve very rapidly in time
while the high momentum piece falls as $t^{-a/3}$ so that for $a = 6$ it fall as $f \sim 1/t^2$ at fixed $E$.
This is however less rapid than $e^{-(t/t_0)^{1/3}}$ for the Boltzmann distribution. This means there can be more radiation at later time.

These considerations suggest that introducing power law tails for quark and gluon distributions can enhance the photon radiation rate and allow the radiation to appear at later times. At later times more flow will have been built up and the produced photon spectra will reflect that.
Hence, this mechanism has the potential to solve the photon flow problem discussed in the introduction.

To understand how the tails might affect the observed distributions of photons, let us consider the radiation from an exponential distribution and compare it to that of a Tsallis distribution. For the following estimate we simply replace the photon distribution by a Tsallis distribution. We will improve on that by replacing quark and gluon distributions by Tsallis distributions and recomputing the photon rate in the following section.

Let us take the formula for thermal radiation Eq.\,(\ref{eq:thermalrate}) ignoring logarithms and constant factors and integrate it over time:
\begin{equation}
  h = \int {{dT} \over T} {T_0^4 \over T^4}  e^{-E_\gamma/T} = \Gamma(4) {T_0^4 \over E_\gamma^{4}}\,.
\end{equation}
This assumes that the temperature of emission $T \sim E_\gamma/4$ is within the range of integration over temperatures.  
Now take a Tsallis distribution
\begin{equation}
 g = \int {{dT} \over T} {T_0^4 \over T^4} (1+E_\gamma/aT)^{-a}\,.
\end{equation}
The stationary phase point of the integral is at
\begin{equation}
 E_\gamma/T = {4a \over {a-4}}\,.
\end{equation}
For $a = 6$, this is $12$, corresponding to a large change in the temperature of emission, which would make for a huge shift in the emission time, which goes as the cube of the temperature.  Clearly such a big shift would move the emission outside of the range of integration over temperature where the QGP assumption is motivated, which will reduce this effect.  Notice also that the value of the integrand at the stationary phase point is $256/9$ which is much greater than the corresponding numerical factor for a Boltzmann distribution.

This is a dramatic example of the effect of tails of distributions.  Our more quantitative results in the following section are fortunately not so dramatic as this example.  
When we use Tsallis distributions for quarks and gluons to compute the rate of emission of photons by scattering, the result is indeed flatter than when using the thermal distributions. 
However, the enhancement in the kinematic region of emission for photons is not as large as in the previous example.  Nevertheless we find significant enhancements of the photon yield and large lengthening of the photon emission times. Another factor that will reduce the effect is the overall lowering of the temperature for a Tsallis distribution relative to a thermal one when keeping the multiplicity fixed.

 We also need to discuss how the Tsallis distribution for quarks and gluons can arise.  In p+p interactions,
 it is thought that Tsallis like tails arise from hadronization of jets.  In the QGP the particles we are describing have not had time to turn into hadrons.  In fact they must be generated from particles with momenta of the order or below the saturation scale.  
A Tsallis distribution is a reasonable guess for the distributions in this region, because the initial production of mini jets naturally leads to power law
tails in addition to softer thermal particles.
  Note that in this picture, the initial typical momentum scales will be proportional to the saturation momentum, and the initial time $t_0 \sim 1/Q_{s}$, so that at least for some time in the evolution we expect that the distributions will scale with the saturation momentum.  If this is the case, then the $N_{\rm part}$ dependence of the initial distribution in the tail of the Tsallis distribution will rise much more rapidly than the $N_{\rm coll}$ assumed for hard particle scattering.  Since the final state distribution of hadrons produced in this kinematic region is not dramatically enhanced, final state interactions, that is jet quenching, must be very important. 

\section{Detailed calculation}\label{sec:detail}
In this section we present a detailed numerical calculation of photon production from a quark gluon plasma, using general distribution functions for quarks and gluons. We compare the numerical results in the thermal limit to the analytic solution in \cite{Baier:1991em,Kapusta:1991qp}, and present the effect of using Tsallis distributions on the photon rate and typical emission time.

\subsection{Photon production rate}
The emission rate of photons with four-momentum $Q=(E_\gamma,\mathbf{q})$
is given by \cite{Baier:1997xc} 
\begin{equation}
 E_\gamma\frac{dR}{d^3 q} = \frac{i}{2(2\pi)^3} {\Pi_{12}}_\mu^\mu (Q) \, ,
 \label{photonrate}
\end{equation}
from the trace of the (12)-element $\Pi_{12}=\Pi^<$ of the
photon-polarization tensor. In a thermal system the rate takes on the form
\cite{Weldon:1983jn,McLerran:1984ay,gk91,Kapusta:1991qp}
\begin{equation}
    E_\gamma\frac{dR}{d^3q}=-\frac{2}{(2\pi)^3}\rm{Im}\Pi_\mu^{\mu}\frac{1}{e^{E_\gamma/T}-1}\,,\label{productionrateselfenergy}
\end{equation}
with the retarded photon self energy $\Pi_{\mu\nu}$. It is valid
to all orders in the strong coupling $\alpha_s$ and to leading
order in $\alpha_{\rm em}$.

When approximating the photon self energy by carrying out a loop
expansion to some finite order Eq.\,(\ref{photonrate}) becomes equivalent to a description in
terms of relativistic kinetic theory.
Generally, expanding the self energy up to $L$ loops is equivalent to
computing the contribution from all reactions of $m$ particles going to $n$ particles, with $m+n\leq L+1$,
with each amplitude calculated to order $g^{L-1}$.
Cutting the one-loop diagram for the photon self energy gives zero for an on-shell photon since the process
$q\bar{q}\to \gamma$ has no phase space. 
Certain cuts of the two-loop diagrams give order $g^2$ corrections to this nonexistent
reaction. Other cuts correspond to the reactions $q\bar{q}\to
g\gamma$, $q g\to q\gamma$ and $\bar{q}g\to\bar{q}\gamma$, the
annihilation and Compton scattering processes.

The contributions of these processes to the photon production rate are
\begin{align}
E_\gamma \frac{d R_{i}}{d^3q} =\mathcal{N}_i \int_{{\bf k}_1}
f_1(\mathbf{k}_1)
        &\int_{{\bf k}_2} f_2(\mathbf{k}_2)
        \int_{{\bf k}_3} (1 \pm f_3(\mathbf{k}_3))\notag\\
       \times(2\pi)^4&\delta^{(4)}(K_1 + K_2 - K_3 - Q) \overline{\left|\mathcal{M}_{i}\right|^2}\,,
\label{eq:rate1}
\end{align}
with $\int_{{\bf k}_i}=\int d^3k_i/(2k_i(2\pi)^3)$. The $f_j$ are
the appropriate distribution functions and there is either a
Bose enhancement or Pauli blocking term
depending on the nature of the strongly interacting particle in
the final state. 
The pre-factor follows from degeneracy factors and 
a sum over the charges of u and d quarks and one obtains $\mathcal{N}_{\rm Compton}=20$ and $\mathcal{N}_{\rm Annihilation}=320/3$.

The matrix elements squared are
\begin{equation}
\overline{\left|\mathcal{M}\right|^2}=-32\pi^2e_q^2\alpha\alpha_s
\frac{1}{6}\left(\frac{t}{s}+\frac{s}{t}\right)\,\label{compmatrix}
\end{equation}
for the Compton process and
\begin{equation}
\overline{\left|\mathcal{M}\right|^2}=32\pi^2e_q^2\alpha\alpha_s\frac{4}{9}\left(\frac{u}{t}+\frac{t}{u}\right)\,\label{annmatrix}
\end{equation}
for the annihilation process.

The matrix elements (\ref{compmatrix}) and (\ref{annmatrix}) have poles at
$t$ and/or $u=0$, which causes infrared divergent production rates (\ref{eq:rate1}).
The screening of these divergences is taken care of by including many body effects in 
a hard loop resummation for soft momentum transfers
$p<p^*$ as done in \cite{Baier:1991em,Kapusta:1991qp}, where $p=|\mathbf{p}|$ and $P = (\omega, \mathbf{p}) = K_1-Q$. In the thermal case and the 
limit $g\rightarrow 0 $ the total rate becomes independent of the cutoff $p^*$ and one obtains Eq.\,(\ref{eq:thermalrate}).

\subsection{Numerical evaluation}

In this work we are interested in the effect of non-thermal quark and gluon distribution functions on photon production.
Even in this case it may be possible to perform a systematic hard loop resummation as done for anisotropic plasmas in 
\cite{Schenke:2006yp}. We leave this study to future work and concentrate here on the effect of the non-thermal distributions on the rates in the region $p>p^*$. 

The general rates for the annihilation and Compton processes in this region read
\bqa E_\gamma \frac{d R_{\rm ann}}{d^3q} &=&
       {16 \frac{5}{9} \alpha_s \alpha}
        \int_{{\bf p}} \frac{ f_q({\bf p}+{\bf q}) }{ \left|{\bf p}+{\bf q}\right| }
        \int_{{\bf k}} \frac{ f_q({\bf k}) }{ k }
        \frac{ 1 + f_g({\bf p}+{\bf k}) }{ \left|{\bf p}+{\bf k}\right| } \nonumber \\
    && \hspace{24mm} \times \,
        \delta(\omega + k - \left|{\bf p}+{\bf k}\right|) \, \Theta(p-p^*) \left[ \frac{u}{t} \right]
        ,
\eqa
and
\bqa E_\gamma \frac{d R_{\rm com}}{d^3q} &=&
       {-16
 \frac{5}{9} \alpha_s \alpha}
        \int_{{\bf p}} \frac{ f_q({\bf p}+{\bf q}) }{ \left|{\bf p}+{\bf q}\right| }
        \int_{{\bf k}} \frac{ f_g({\bf k}) }{ k }
        \frac{ 1 - f_q({\bf p}+{\bf k}) }{ \left|{\bf p}+{\bf k}\right| } \nonumber \\
    && \hspace{24mm} \times \,
        \delta(\omega + k - \left|{\bf p}+{\bf k}\right|) \, \Theta(p-p^*) \left[ \frac{s}{t} + \frac{t}{s} \right]
        ,
\eqa
where we relabeled $\mathbf{k}_2$ to $\mathbf{k}$ and combined the two terms in the matrix element for the annihilation process assuming 
that the distribution functions for quarks and anti-quarks are the same.

We simplify these expressions following \cite{Schenke:2006yp} and evaluate them using Monte Carlo integration.

\begin{figure}[htb]
  \begin{center}
    \includegraphics[width=0.8\textwidth]{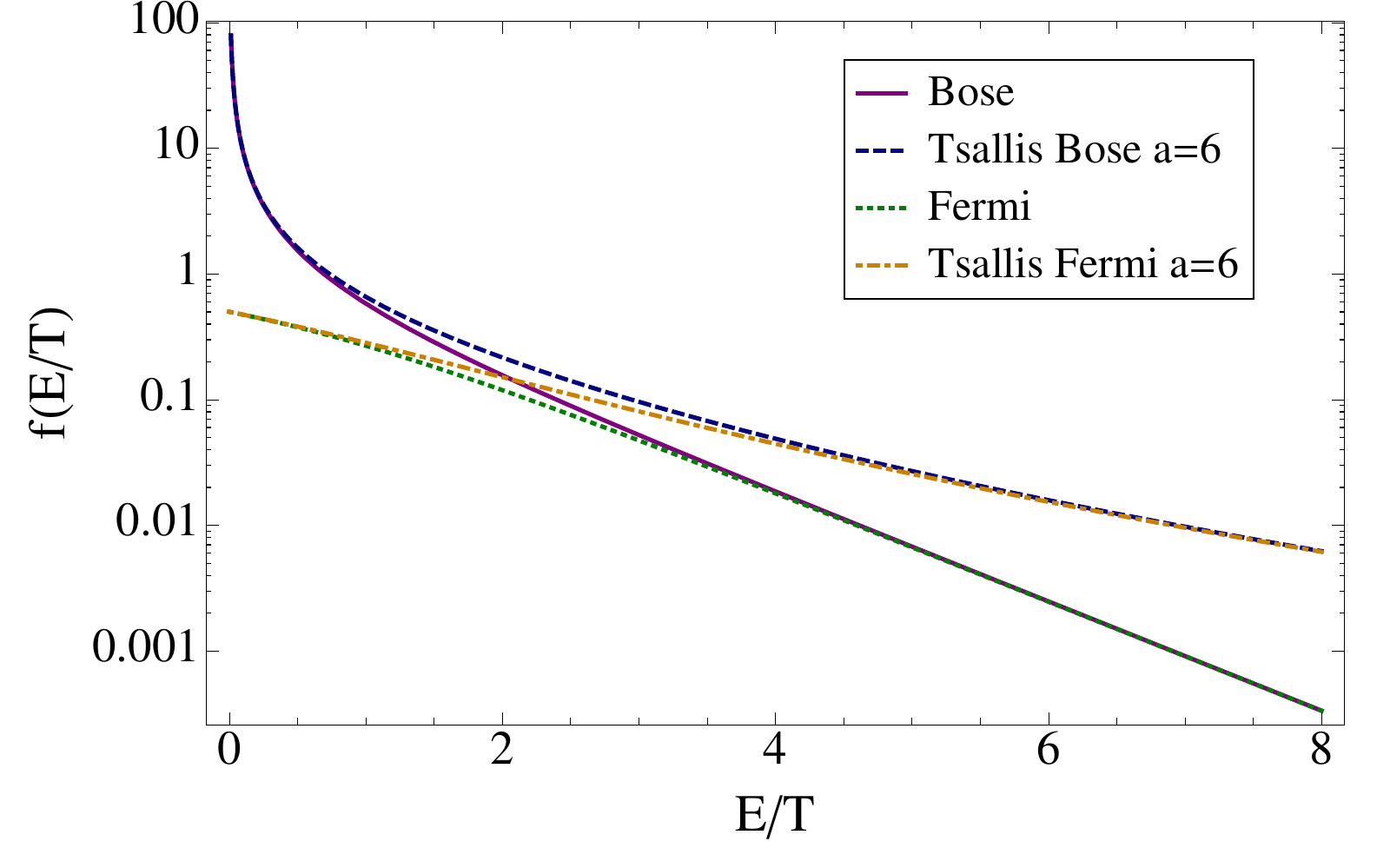}
    \caption{Comparison of thermal and Tsallis distribution functions with $a=6$. $E/T$ is the energy of the quark or gluon scaled by the temperature $T$.}
    \label{fig:distributions}
  \end{center}
\end{figure}

To study the effect of non-thermal tails in the distribution functions, we compare the thermal rates obtained by using Bose-Einstein and 
Fermi-Dirac distributions for gluons and quarks, respectively, with rates where those distributions were replaced by their Tsallis counterparts:
\begin{align}
  f_q^{\rm a} (E_\gamma) &= \left[\left(1+\frac{E_\gamma}{a T}\right)^a+1\right]^{-1}\,,\label{eq:TsallisQ}\\
  f_g^{\rm a} (E_\gamma) &= \left[\left(1+\frac{E_\gamma}{a T}\right)^a-1\right]^{-1}\,,\label{eq:TsallisG}
\end{align}
where the parameter $a$ characterizes the power law tail of the distribution.
We show a comparison of the thermal rates to those defined in (\ref{eq:TsallisQ}) and (\ref{eq:TsallisG}) in Fig.\,\ref{fig:distributions}.

\begin{figure}[htb]
  \begin{center}
    \includegraphics[width=0.8\textwidth]{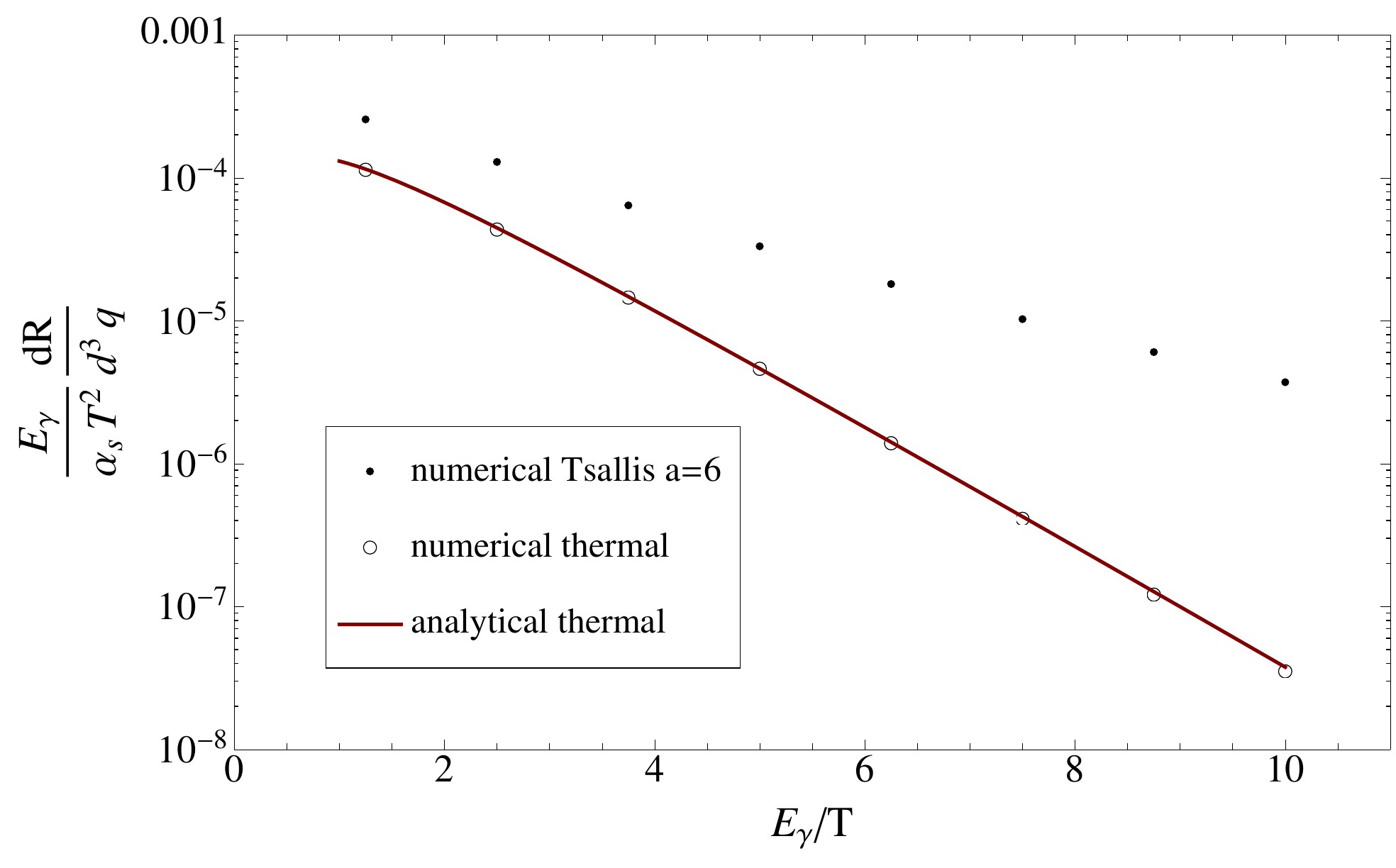}
    \caption{Comparison of thermal and Tsallis ($a=6$) rates for $p^*\approx 0.56$. 
      Also shown is the analytic result for the hard part of the thermal rate from \cite{Baier:1991em}, which agrees perfectly with our numerical result. }
    \label{fig:rates}
  \end{center}
\end{figure}

We present a comparison of the thermal photon rates with those obtained using Tsallis distributions with parameter $a=6$ for quarks and gluons in 
Fig.\,\ref{fig:rates}. We chose the infrared cutoff $p^*=  10^{-0.25} \approx 0.56$ to lie in the region where the soft and hard contributions are approximately equal.
This means the full rate (sum of the soft HTL resummed and the hard contribution) is approximately a factor of 2 larger.

\begin{figure}[htb]
  \begin{center}
    \includegraphics[width=0.8\textwidth]{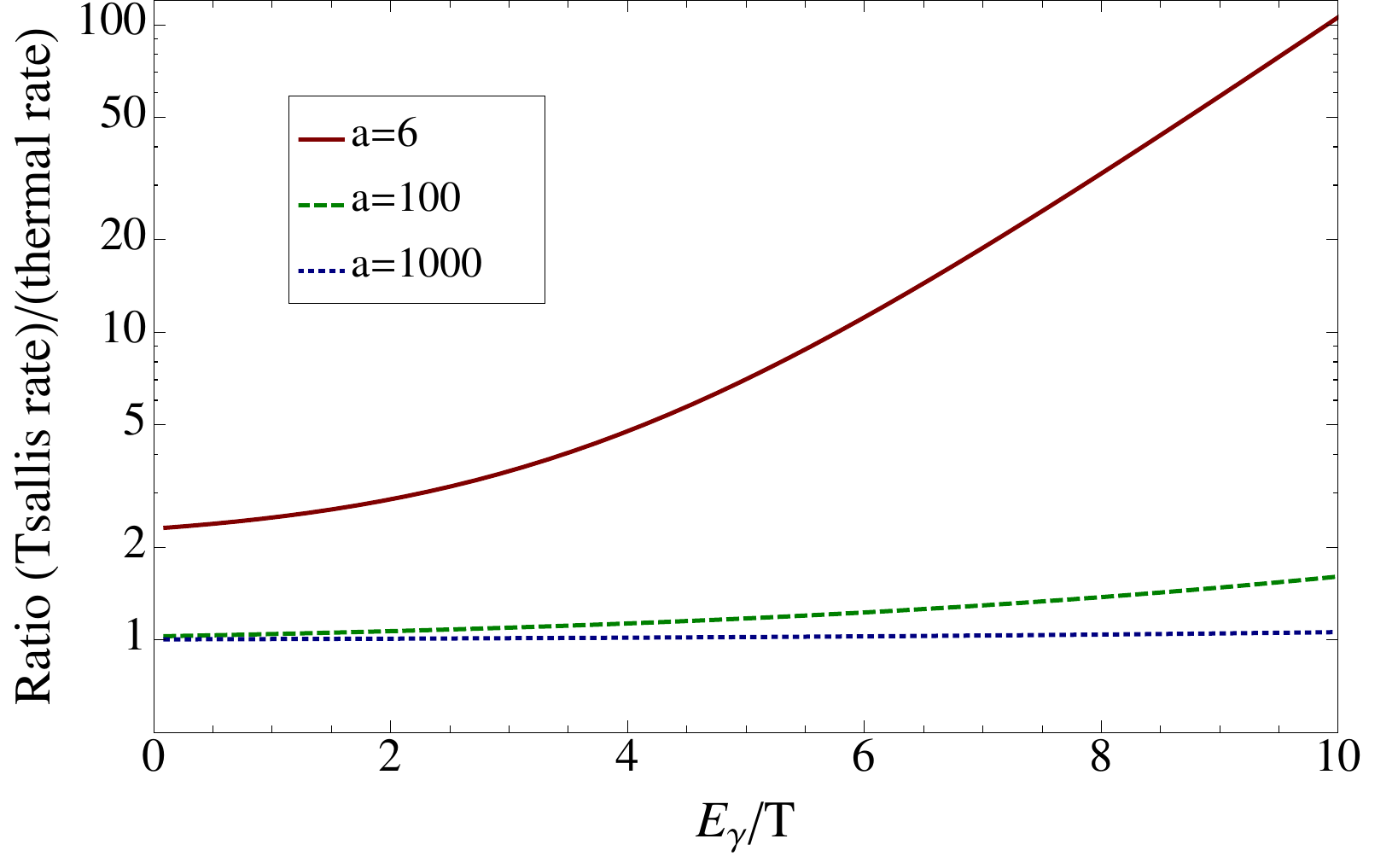}
    \caption{Exponential fits to the ratio of the Tsallis to the thermal rate for $p^*\approx 0.56$ and different $a$. Only for very large $a$ does
the Tsallis rate approach the thermal limit.}
    \label{fig:ratios}
  \end{center}
\end{figure}

One can see that the effect of the tails in quark and gluon distributions is significant. The photon rate becomes larger and harder.
To show the effect more clearly we plot the ratio of the Tsallis to the thermal rate for different parameters $a$ in Fig.\,\ref{fig:ratios}.
In fact, the curves shown in Fig.\,\ref{fig:ratios} are exponential fits to the ratios. We find that using Tsallis distributions with parameter $a=6$ leads to almost exponential photon spectra but with an effective temperature twice as large as the temperature scale used in the calculation.
One can also see in Fig.\,\ref{fig:ratios} that the Tsallis result approaches the thermal result very slowly. Only for parameters $a\approx 1000$ is the difference in rates negligible.

\begin{figure}[htb]
  \begin{center}
    \includegraphics[width=0.8\textwidth]{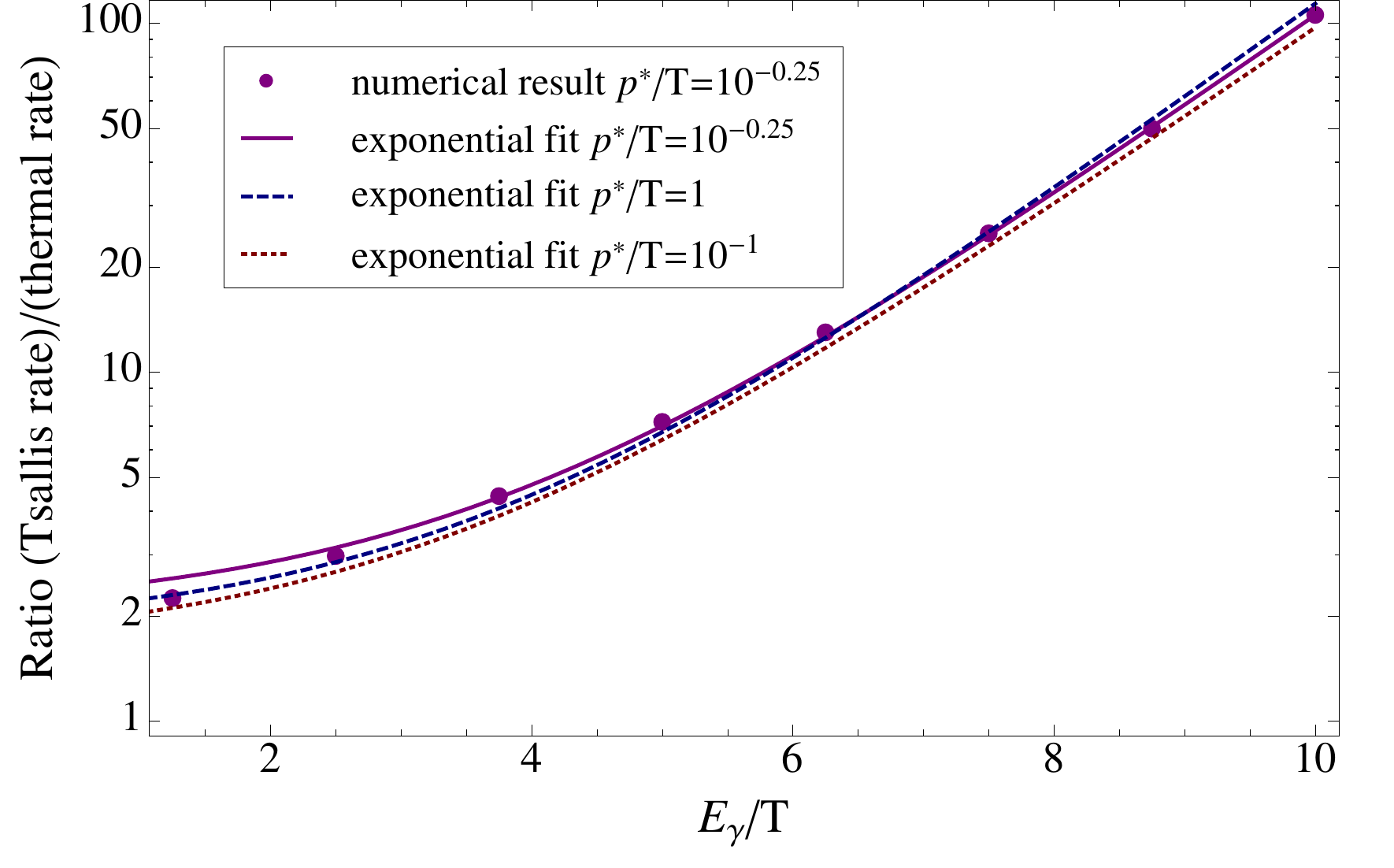}
    \caption{Dependence of the ratio of Tsallis to thermal rate on the infrared cutoff $p^*$ for $a=6$.}
    \label{fig:cutoff}
  \end{center}
\end{figure}

Fig.\,\ref{fig:cutoff} shows that the ratios of Tsallis to thermal rates are almost independent of the infrared cutoff $p^*$.
This motivates us to modify the full thermal rate obtained using HTL resummation by the same factor to compute the total photon yield.
Explicitly, for $a=6$ we find that the ratio is well approximated by 
\begin{equation}\label{eq:rfactor}
 r(E_\gamma/T) = \frac{\left(\frac{d R^{\rm Tsallis}}{d^3q}\right)}{\left(\frac{d R^{\rm thermal}}{d^3q}\right)}  = 2.06 +  0.238 \exp(0.608\,E_\gamma/T)\,.
\end{equation}

\begin{figure}[htb]
  \begin{center}
    \includegraphics[width=0.8\textwidth]{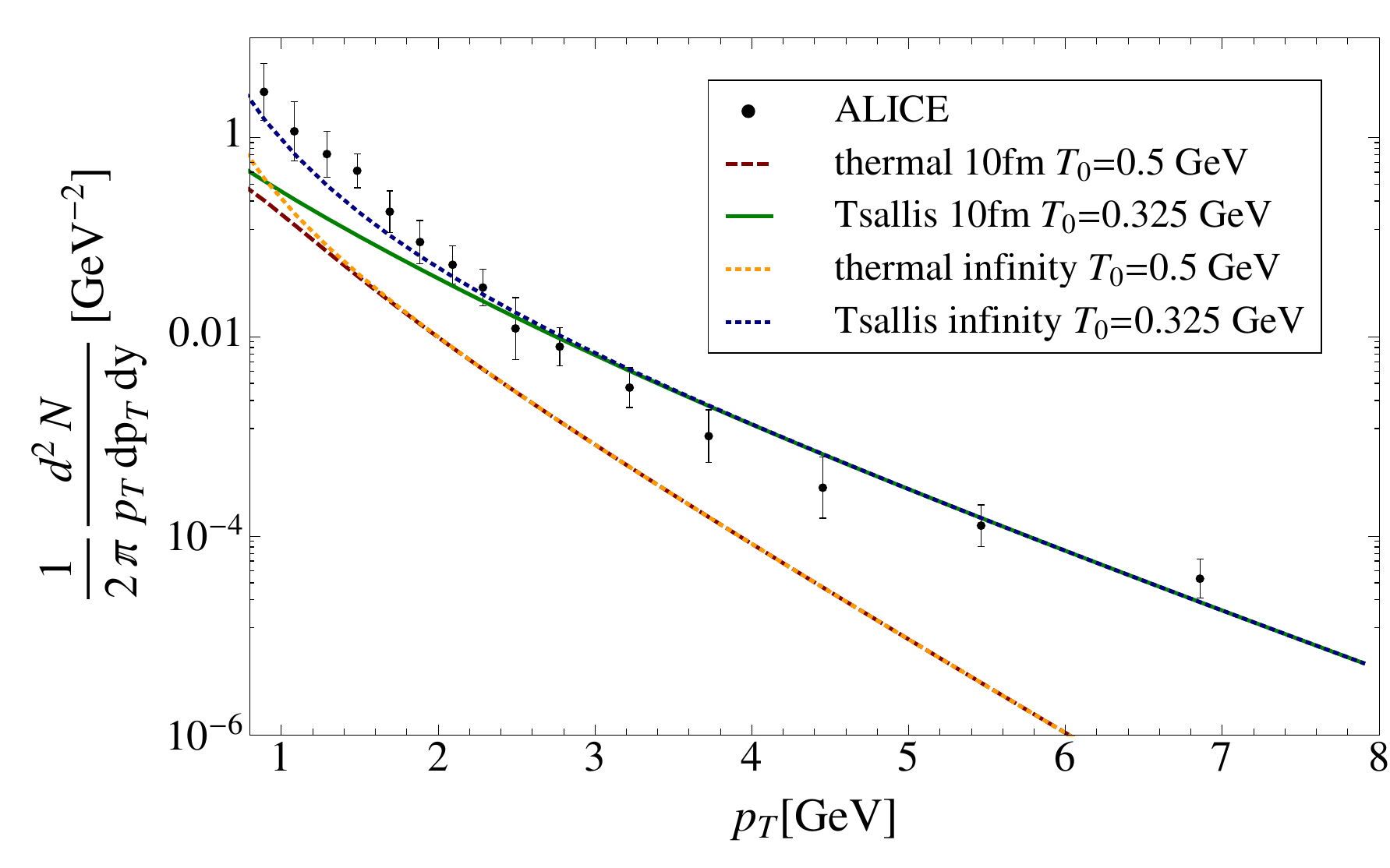}
    \caption{Photon yields after emitting for $10\,{\rm fm}/c$ and infinity in the thermal case with $T_0=500\,{\rm MeV}$ and using Tsallis quark and gluon distributions with $T_0=325\,{\rm MeV}$ and $a=6$. The temperatures were chosen such that the number of quarks and gluons are the same in both cases. We compare to ALICE data for 0-40\% central collisions from \cite{Wilde:2012wc}.}
    \label{fig:yieldsTdependence}
  \end{center}
\end{figure}

To compute the produced photons from a one-dimensionally expanding quark gluon plasma, we integrate the photon rate using a time dependent temperature $T(t) = T_0 (t_0/t)^{1/3}$, and set $\alpha_s=1/3$. The transverse size of the system is fixed to $A_T=\pi R^2$ with $R=6.6\,{\rm fm}$, the radius of a lead nucleus.
In the thermal case, we choose $T_0=500\,{\rm MeV}$, starting at time $t_0 = 1/T_0$. This value for $T_0$ is extracted from numerical fluid dynamic simulation results at LHC energies that produce hadron spectra and flow harmonics in agreement with experimental data \cite{Schenke:2011tv}.
In case of the Tsallis distribution we multiply the thermal rate (\ref{eq:thermalrate}) by the factor $r(E_\gamma/T(t))$ given in (\ref{eq:rfactor}), and re-scale the initial temperature to ensure the same total number of quarks and gluons as in the thermal case.
Assuming Boltzmann distributions, this is achieved when re-scaling the initial temperature in the Tsallis case by a factor 
\begin{equation}
 N_T(a)=\left(\frac{\Gamma(a)}{a^3 \Gamma(a-3)}\right)^{1/3},\
\end{equation}
which is $N_T\approx 0.65$ for $a=6$. In case of the Bose or Fermi distributions, we get slightly different results of $N_T(6) = 0.67$, and $N_T(6) = 0.64$, respectively. For simplicity, we will use the factor obtained assuming Boltzmann distributions in this work.
The obtained photon yields are shown in Fig.\,\ref{fig:yieldsTdependence}. 

\subsection{Photon emission times}
We now compute the typical emission times for photons of different energies according to
\begin{equation}
 \langle t \rangle_{\rm th} = \frac{  \int_{t_0}^{t_{\rm max}}t^2 dt\, (dR(t)/dy~p_Tdp_T) }{  \int_{t_0}^{t_{\rm max}} t dt\, (dR(t)/dy~p_Tdp_T)}\,.
\end{equation}
Results using $t_{\rm max}=10\,{\rm fm}/c$ are presented in Table\,\ref{tab:meanTimes}. We find that the harder tails in the Tsallis distributions lead to significantly longer 
mean emission times for photons. This effect can potentially delay photon emission so much that photons will be produced from a medium that has
most of its anisotropic flow built up, leading to an increased photon elliptic flow. Detailed calculations of this using state of the art
 3+1 dimensional hydrodynamics is the subject of future work.

\begin{table}
\begin{center}
    \begin{tabular}{|c|c|c|c|c|}
    \hline
    $\langle t \rangle\,[{\rm fm}/c]$ & th. $(T_0=500\,{\rm MeV})$ & Tsallis $(T_0=325\,{\rm MeV})$ \\\hline
    $p_T=1\,{\rm GeV}$    &  3.9   &  4.6   \\\hline
    $p_T=2\,{\rm GeV}$    &  1.7    &  3.5    \\\hline
    $p_T=3\,{\rm GeV}$    &  0.9    &  2.5    \\\hline
    \end{tabular}
\end{center}
\caption{Mean photon emission time using thermal and Tsallis rates and $t_{\rm max}=10\,{\rm fm}/c$. The harder tails of the Tsallis distribution lead to significantly longer emission times compared to the thermal case, even when adjusting the initial temperatures to ensure equal numbers of partons in the system. \label{tab:meanTimes}}
\end{table}

\begin{figure}[htb]
  \begin{center}
    \includegraphics[width=0.8\textwidth]{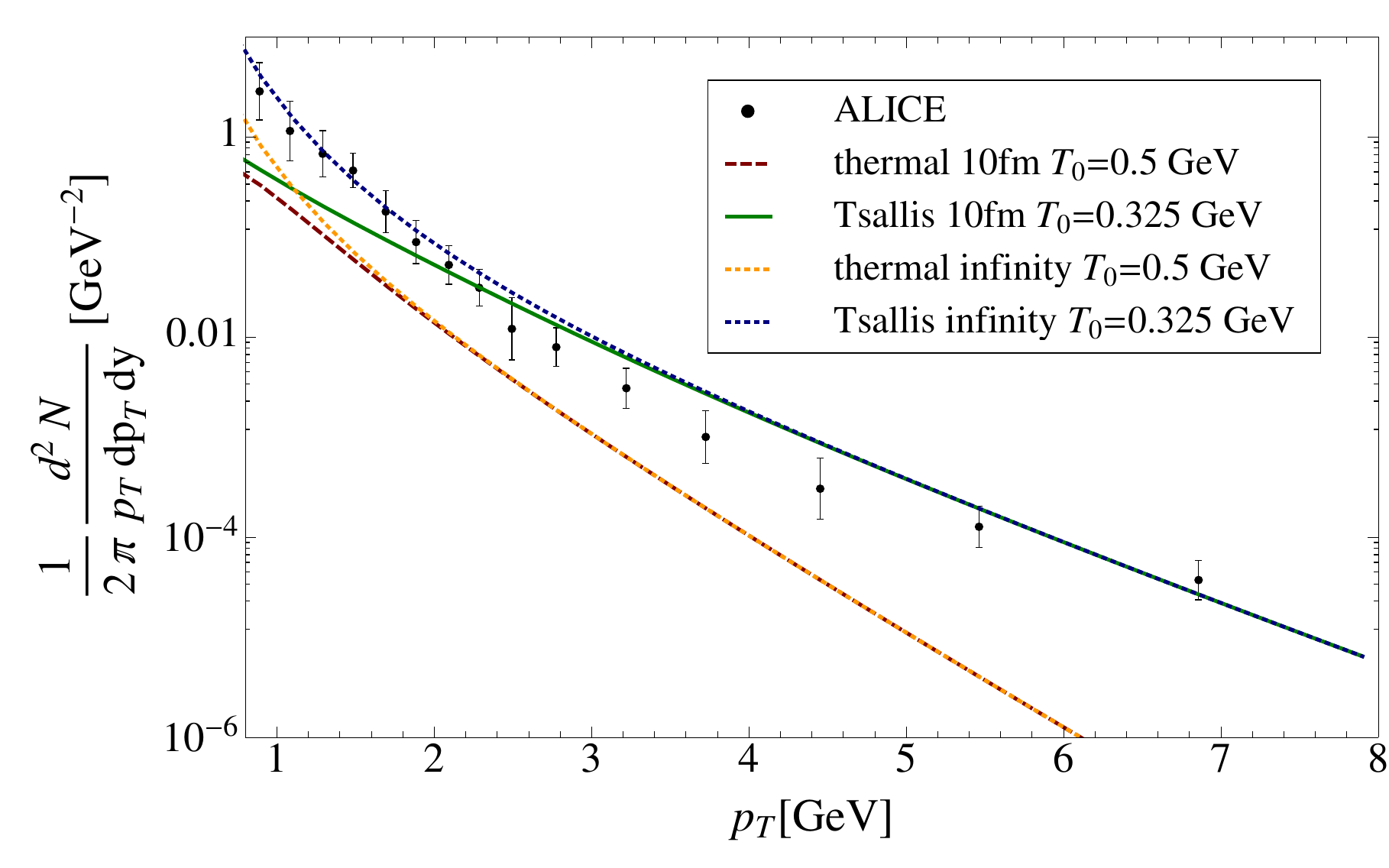}
    \caption{Same as Fig.\,\ref{fig:yieldsTdependence} but using a modified time dependence of the temperature $T(t)=T_0(t_0/t)^{9/30}$.
      We compare to ALICE data for 0-40\% central collisions from \cite{Wilde:2012wc}.}
    \label{fig:yieldsTdependence930}
  \end{center}
\end{figure}

Finally, we show that the slope of the photon spectrum is quite sensitive to the time dependence of the temperature evolution.
Decreasing the power $1/3$ by 10\%, such that $T(t)=T_0(t_0/t)^{9/30}$, leads to a steepening of the photon spectra as shown in Fig.\,\ref{fig:yieldsTdependence930}. In particular, this leads to better agreement with the experimental data in the low momentum region.
Note that here we have not fixed the number of quarks and gluons to be the same as in the case of the usual 1+1 dimensional expansion, since we only want to emphasize the effect on the shape of the spectrum.

\section{Summary and Conclusions}\label{sec:conc}

We have shown that a relatively small modification of thermal distributions to include power law tails results in dramatic modifications of the computed thermal photon yields and the times of emission.  This modification has the correct properties to solve the problems with 
a theoretical description of photons from heavy in collisions as encountered for RHIC and LHC energies.
To establish this explicitly, one must incorporate the modifications introduced in this work into a realistic 3+1 dimensional hydrodynamic simulation of the collision.

 To compute the power law tails from first principles may involve either a proper classical field computation, or cascade simulations where power law tails are dynamically generated.  Such computations would allow one to assess the reasonability of our assumptions that the tails scale with the same momentum $\sim T$, as does the center of the distribution, and that the power is close to $a = 6$.
 
The emission times we find are possibly long enough for part of the emission to occur in the hadron gas phase,
where one may also expect power law tails of hadron distributions.
It is however conceivable that in a strongly interacting medium such tails may be best described by partonic degrees of freedom as well. 
Such tails might have the property of geometric scaling since they arise from high energy processes,
and thus would be consistent with the observed scaling behavior.
 
Given the $A$ dependence of the distribution functions in various transverse momentum regimes, 
our results suggest that the primeval distribution of quarks and gluons in heavy ion collisions
has a significant Cronin enhancement at intermediate transverse momentum values.
This should have implications for jet quenching computations. 

 \section{Acknowledgments}
We thank Charles Gale and Krzysztof Redlich for very useful comments.
The authors are supported under Department of Energy Contract No. DE-SC0012704. BPS acknowledges a DOE Office of Science Early Career Award.

\bibliographystyle{elsarticle-num-names}
\bibliography{spires}

\end{document}